\documentclass[conference]{IEEEtran}
\IEEEoverridecommandlockouts

\usepackage{cite}
\usepackage{amsmath,amssymb,amsfonts}
\usepackage{graphicx}
\usepackage{textcomp}
\usepackage{xcolor}
\usepackage{float} 
\usepackage[bookmarks=false]{hyperref} 
\usepackage{amsmath}
\usepackage{array}
\usepackage{graphicx}
\usepackage{tabularx} 
 
\usepackage{caption} 
\usepackage{hyperref} %
\usepackage{etoolbox} 
\usepackage{xcolor}
\usepackage{amsmath}
\usepackage{algorithm}
\usepackage{algpseudocode}
\usepackage{dsfont}

    \def\BibTeX{{\rm B\kern-.05em{\sc i\kern-.025em b}\kern-.08em
    T\kern-.1667em\lower.7ex\hbox{E}\kern-.125emX}}
\begin{document}

\title{Adaptive DRL for IRS Mirror Orientation in Dynamic OWC Networks\\
}

\author{\IEEEauthorblockN{Ahrar N. Hamad, Ahmad Adnan Qidan, Taisir E.H. El-Gorashi and Jaafar M. H. Elmirghani, }
\IEEEauthorblockA{\textit{ Department of Engineering, King’s College London, London, United Kingdom,} \\
\textit
\{{ahrar.hamad, ahmad.qidan, taisir.elgorashi, jaafar.elmirghani}\}@kcl.ac.uk}
\thanks{This work has been supported by the Engineering and Physical Science Research Council (EPSRC), by the INTERNET project under Grant EP/H040536/1, by the STAR project under Grant EP/K016873/1,  by the TOWS project under Grant EP/S016570/1, and by the TITAN project under Grant EP/X04047X/2. All data are provided in full in the results section of this paper.}
}
\maketitle


\begin{abstract} Intelligent reflecting surfaces (IRSs) have emerged as a promising solution to mitigate line-of-sight (LoS) blockages and enhance signal coverage in optical wireless communication (OWC) systems with minimal additional power. In this work, we consider a mirror-based IRS to assist a dynamic indoor visible light communication (VLC) environment. We formulate an optimization problem that aims to maximize the sum rate by adjusting the orientation of the IRS mirrors. To enable real-time adaptability, the problem is modelled as a Markov decision process (MDP), and a deep reinforcement learning (DRL) algorithm is developed based on the deterministic policy gradient for real-time mirror-based IRS optimization in dynamic VLC networks. The proposed DRL is employed to optimize mirror orientation toward mobile users under blockage and mobility constraints. Simulation results demonstrate that our proposed DRL algorithm outperforms the conventional deep Q-learning (DQL) algorithm and achieves substantial improvements in sum rate compared to random-orientation IRS configurations. 

\end{abstract}
\vspace{+2pt}
\begin{IEEEkeywords}
 Intelligent reflecting surface (IRS), optical wireless communication (OWC),  visible light communication (VLC), reinforcement learning (RL).
\end{IEEEkeywords}

\section{Introduction}\label{Sec:Intro}
Optical wireless communication (OWC) is envisioned as a promising technology to meet green communication requirements in sixth-generation (6G) networks,  as it offers higher bandwidth, improved energy efficiency, and reduced electromagnetic pollution compared to conventional radio frequency (RF) systems \cite{Q9521837}.
Among various OWC technologies, visible light communication (VLC) has gained significant attention due to its ability to provide both lighting and data transmission in indoor environments. However, VLC systems face key challenges, particularly in maintaining reliable coverage. This is because the coverage area of an optical access point (AP) is confined, and line-of-sight (LoS) signals are easily obstructed by physical objects  \cite{sun2021sum}. 

The integration of intelligent reflecting surfaces (IRS) offers a promising approach to improve connectivity in indoor VLC systems without consuming much power \cite{9838853}. IRS extend coverage by reflecting non-line-of-sight (NLoS) optical signals towards useres, thereby enhancing user experience. Two primary technologies are considered for IRS implementation: mirror arrays and metasurfaces. A mirror array consists of small, passive, and individually controllable mirrors, while a metasurface comprises sub-wavelength elements (meta-atoms) arranged in a planar array. In \cite{abdelhady2021visible}, the performance of these IRS types in VLC systems was analyzed. The results indicate that the effectiveness of power focusing depends on the number of IRS reflecting elements as well as the dimensions of the source and reflector. Steerable IRSs offer flexibility by dynamically controlling and directing reflected signals toward mobile users in rapidly changing environments. 

Recent studies have proposed various algorithms to optimize IRS element orientation in VLC networks. For instance, an efficient sine-cosine algorithm was introduced in \cite{aboagye2021intelligent} to determine the optimal IRS configuration for establishing a reliable NLoS link. In \cite{9500409}, a particle swarm optimization algorithm was developed for an IRS-assisted VLC system to maximize secrecy by identifying optimal mirror orientations. However, conventional optimization algorithms often face limitations in solving complex problems in real-time as they rely on static models and assumptions and lack adaptability to dynamic conditions \cite{elgamal2021reinforcement}. To address these limitations, reinforcement learning (RL) has emerged as a powerful tool that supports adaptive and intelligent decision-making \cite{sutton2018reinforcement}.

RL can derive policies by interacting with the environment, making it well-suited for dynamic wireless network scenarios. 
In \cite{hamad2024reinforcement}, conventional RL algorithms such as Q-learning and state-action reward-state-action (SARSA) were employed to improve data rates in indoor IRS-aided OWC systems. However, these methods suffer from memory inefficiency due to their reliance on a Q-table, which becomes computationally infeasible in large state-action spaces.  To overcome this, deep Q-learning (DQL) was proposed in \cite{al2023deep} to manage VLC networks and enhance spectral efficiency. DQL leverages experience replay and target networks to improve sample efficiency and training stability. Nevertheless, it still struggles with continuous action spaces, which are common in practical scenarios. To address this, a deep reinforcement learning (DRL) algorithm was proposed in \cite{9784887}, combining the strengths of deterministic policy gradients and DQL to effectively manage continuous action spaces, overcoming the limitations of DQL-based methods.


\begin{figure*}[t!]
    \centering
    \includegraphics[width=0.8\textwidth]{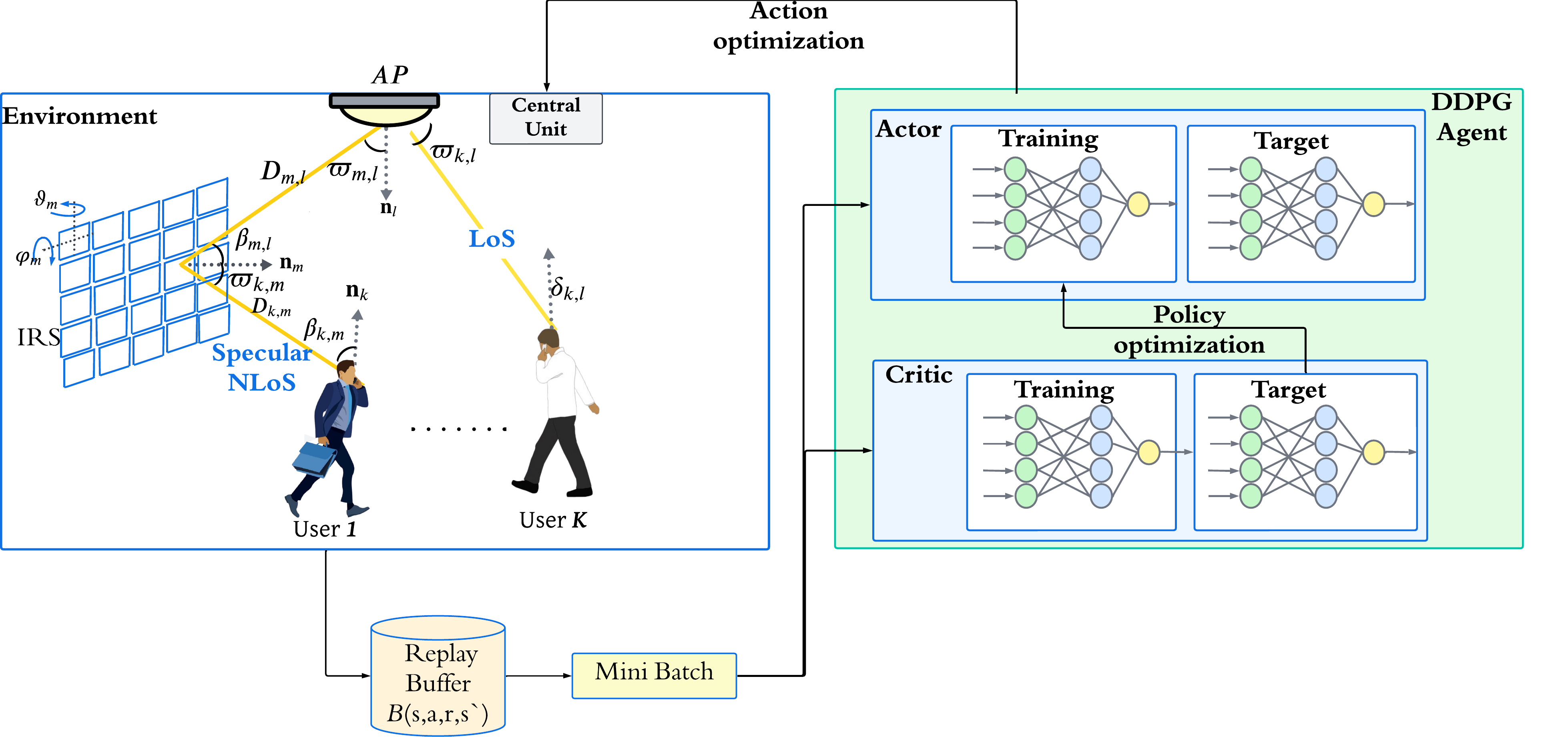}
    \caption{IRS-assisted VLC system model with the DRL framework.} 
    \label{fig:System_model}
    \vspace{-10pt}
\end{figure*}

In contrast to the prior works that optimize IRS configuration as an association matrix or assume static users, we develop a DRL-based algorithm based on a deterministic policy to jointly optimize IRS mirror orientations for multi users while ensuring quality of service (QoS) constraints for each user in indoor environments. The optimisation problem is formulated as a Markov decision process (MDP) and solved using the DRL framework \cite{9784887}. The algorithm is trained offline and tested in real-time to improve the QoS for multiple mobile users traveling at varying velocities. We evaluate system performance considering realistic dynamic conditions, including blockage, mobility and channel variations. Our results demonstrate that the proposed DRL algorithm significantly enhances user experience compared to DQL and conventional algorithms.

The rest of this paper is structured as follows: Section \ref{Sec:System_model} presents the system model. Section \ref{Sec:ProbForm} formulates the optimization problem. Section \ref{Sec:RL} explains the application of RL algorithms to solve the problem. Section \ref{Sec:Results} discusses the simulation results. Finally, Section \ref{Sec:Conc} concludes the paper.

\section{System Model}\label{Sec:System_model}

We consider a downlink IRS-assisted VLC system, as shown in Fig. \ref{fig:System_model}. An array of light-emitting diode (LEDs) is placed on the ceiling as an optical AP, $l$, for communication and illumination. On the communication floor, a number of mobile users, $\mathcal{K}=[1,\dots, K]$, move at different velocities. Each user is equipped with a multi-branch angle diversity receiver (ADR) assumed to be upward-oriented as in \cite{eldeeb2018interference}. Furthermore, an IRS mirror array is mounted on one of the walls and consists of $\mathcal{M}=[1,\dots, M_{ij}]$ mirrors. The area of each mirror is $dA_m = h_m \times w_m$, where $h_m$ and $w_m$ represent the height and width, respectively. A space $q_m$ is considered between the mirrors. In addition, the orientation of the rotational mirror can be controlled by two angles: the roll angle, $\varphi_m$, and the yaw angle, $\vartheta_m$, via the micro-electro-mechanical systems (MEMS) technology \cite{10080950}. This rotation allows the mirror to steer the reflected signal from the AP in the desired direction toward the users, as shown in Fig. \ref{fig:System_model}. The network is managed through a central unit (CU), which gathers global information through a WiFi AP and manages resource allocation in real time.

\subsection{VLC channel model}\label{AA}
\textit{1) LoS channel gain:}
The channel gain of the direct LoS between user $k$ and the optical AP is given by

\begin{equation}
H_{k,l}^{\text{LoS}} = 
\begin{cases} 
\frac{(n+1) A_r \cos^n(\varpi_{k,l}) \cos(\delta_{k,l})}{2 \pi D_{k,l}^2}, & 0 \leq \delta_{k,l} \leq \psi_c \\
\quad \quad 0, & \delta_{k,l} > \psi_c
\end{cases}
\end{equation}

where \( n = -\ln(2)/\ln(\cos(\phi_{1/2})) \) is the order of Lambertian emission, which is based on the LED half-power semi-angle \(\phi_{1/2}\). \( A_r \) is the area of the detector,
$\varpi_{k,l}$ and $\delta_{k,l}$ are the irradiance angle and incidence angle, respectively. \( D_{k,l} \) is the LoS transmission distance between AP \( l \) and user \( k \). Note that, the incidence angle, \( \delta_{k,l} \), must be within a range from 0 to the acceptance semi-angle of the concentrator, \( \psi_c \), to guarantee that the LoS signal is detected by the receiver, otherwise, no signal is received, i.e., $h_{k,l}^{\text{LoS}}$ = 0.

\textit{2) IRS NLoS channel gain:}
The channel gain of the indirect communication path is reflected by the mirrors of the IRS with complete specular reflection. Note that, the interference between the specular IRS signals can be ignored \cite{sun2021sum}. The received channel gain by user $k$ from  mirror \( m \in m_i \times m_j\), reflecting the signal of the optical AP is given by

\vspace{-7pt}
\begin{equation} \label{IRSCH}
H_{k,m,l}^{\text{IRS}} = 
\begin{cases} 
\frac{(n+1) \rho_m A_r dA_m \cos^n(\varpi_{m,l}) \cos(\beta_{m,l}) \cos(\varpi_{k,m}) \cos(\beta_{k,m})}{2\pi^2 (D_{m,l})^2 (D_{k,m})^2} \\
\quad \quad , 0 \leq \beta_{k,m} \leq \psi_c \\
\quad 0, ~~~~~~~\beta_{k,m} > \psi_c,
\end{cases}
\end{equation}

where \( \rho_m \) is the mirror reflection coefficient and  \( \varpi_{m,l} \) is the irradiance angle from the optical AP to mirror \( m \), $\beta_{m,l}$ is the incidence angle from AP $l$ to mirror $m$, $\varpi_{k,m}$ is the irradiance angle from mirror $m$ towards user $k$, \( \beta_{k,m} \) is the incidence angle of the signal reflected from mirror \( m \) to user \( k \), \( D_{m,l} \) is the distance between the optical AP and mirror \( m \), and \( D_{k,m} \) is the distance between mirror \( m \) and user \( k \). %

In this work, the IRS mirrors are rotational to allow for dynamic adjustments of reflection angles. The orientation of the IRS mirrors must be controlled by yaw $\varphi_m$ and roll $\vartheta_m$ angles. Their impact on the user channel gain can
be captured through the cosine $\cos(\beta_{k,m})$ as follows \cite{aboagye2022design} 

\begin{equation}
\label{IRSo}
\begin{aligned}
\cos(\beta_{k,m}) = & \bigg(\frac{x_m - x_k}{D_{k,m}} \bigg) \sin(\varphi_m) \cos(\vartheta_m)  
+ \bigg(\frac{y_m - y_k}{D_{k,m}} \bigg)\\
&\cos(\varphi_m) \cos(\vartheta_m) 
+ \bigg(\frac{z_m - z_k}{D_{k,m}} \bigg) \sin(\vartheta_m).
\end{aligned}
\end{equation} 
\noindent where $(x_m, y_m, z_m)$ represent the coordinates vector of the mirror $m$.

\subsection{ User Mobility and Blockage}\label{AAA}
We employ RWP to model the movement of humans within the indoor space and use the Matern hard-core process (MHCP) to model static obstacles, such as furniture or static humans, in the environment \cite{jacob2013fundamental}.
 
In RWP, each user moves at a uniform velocity in the range $[v_{min}, v_{max}]$. Furthermore, the user randomly selects a destination waypoint within the room according to a uniform distribution. The user moves toward that point with another velocity along a straight line.  
When arriving at the waypoint, the user pauses for a randomly determined interval before selecting a new waypoint. Once the user reaches their destination, the process repeats \cite{9314907}. As users move within the environment, their position distribution gradually approaches a steady-state distribution. The stationary distribution of the user locations in an area of \( a \times a \) size is given by 

\begin{equation}
f_X(\mathbf{x}) = f_{XY}(x, y) = \frac{36}{a^6} \left( x^2 - \frac{a^2}{4} \right) \left( y^2 - \frac{a^2}{4} \right),
\end{equation}

\noindent where \( -\frac{a}{2} \leq x \leq \frac{a}{2} \) and \( -\frac{a}{2} \leq y \leq \frac{a}{2} \) represent the coordinates of each node within the square area, and \( \mathbf{x} = [x, y] \) denotes the position vector. 

The MHCP model distributes stationary human blockages within the  {which reflects the realistic blockage patterns for VLC environment.In MHCP, two or more points are separated by a minimum distance so that they do not overlap. Moreover, the number of blockages is proportional to the area of the room \cite{9314907}.

The blockage is modeled as a cylinder with a height of $h_{\mathcal{B}_l}$ and a diameter of $D_{\mathcal{B}_l}$. When a blockage of height $h_{\mathcal{B}_l}$ is positioned between the LED and the receiver at distances $d_{\mathcal{B}_l}$ and $d_L$, respectively, it creates a shadow region. Using geometric principles as shown in \ref{fig:Blockage}, the shadow length $(d_L - d_{\mathcal{B}_l})$ can be calculated as \cite{jacob2013fundamental}

\begin{equation}
  d_L - d_{\mathcal{B}_l} = \frac{(h_{\mathcal{B}_l} d_L)}{h_l},
\end{equation}
This shadow area is created as a rectangular shape with length $(d_L - d_{\mathcal{B}_l})$ and width equal to the diameter of the blockage.

\subsection{User data rate}
The signal-to-interference-plus-noise ratio (SINR), $\textit{ $ \Gamma $}_k$, of  user $k$ can be expressed as
\begin{equation} 
\Gamma_k = \frac{ R_0^2 P \left(\mathds{1} H_{k,l}^{\text{LoS}} + \sum_{m \in \mathcal{M}} H_{k,m,l}^{\text{IRS}}  \right)^2 }{ I_k^2 + \sigma_t^2 }, 
\label{snr}
\end{equation}

where $R_0$ is the optical-electric responsivity coefficient of the photodetector, $P$ is the power to serve user $k$, $\mathds{1}(.)$ is a binary indicator function, its value depends on the severity of the blockage, $I_k^2 $ is the residual interference, and $\sigma_t^2$ is the summation of preamplifier noise, shot noise, and background noise. From \eqref{snr}, The user data rate is given by \cite{wang2013tight}
\begin{equation}
R_k = \frac{B_w}{K} \log_2 \left(1 + \frac{e}{2\pi} \Gamma_k \right),
\end{equation}
where $B_w$ represents the system modulation bandwidth and ${e}$ accounts for the Euler number. In our network, the bandwidth is divided by $K$ to avoid multi-user interference.


\section{Problem Formulation}\label{Sec:ProbForm}
\vspace{-2pt}

We now formulate an optimization problem  to find the optimal IRS orientations towards the users as follows

\begin{equation}
\textbf{P:} \quad
\max_{ \boldsymbol{\varphi}, \boldsymbol{\vartheta}} \sum_{k \in \mathcal{K}}  R_k, 
\label{eq:P}
\end{equation}
\begin{subequations}
    \renewcommand{\theequation}{8\alph{equation}}
    \begin{equation}
        \text{s.t.}\quad R_k \geq R_{\text{min},k}, \quad \forall k \in \mathcal{K},
        \label{eq:cons1}
    \end{equation}
    \begin{equation}
        -\frac{\pi}{2} \leq {\varphi}_{m} \leq \frac{\pi}{2}, \quad \forall m \in \mathcal{M}, 
        \label{cons:2}
    \end{equation}
    \begin{equation}
        -\frac{\pi}{2} \leq {\vartheta}_{m} \leq \frac{\pi}{2}, \quad \forall m \in \mathcal{M}.
        \label{cons:3}
    \end{equation}
    \label{Eq:op}
\end{subequations}

\setcounter{equation}{8} 

\begin{figure}[t!]
    \centering
    \includegraphics[width=0.5\textwidth]{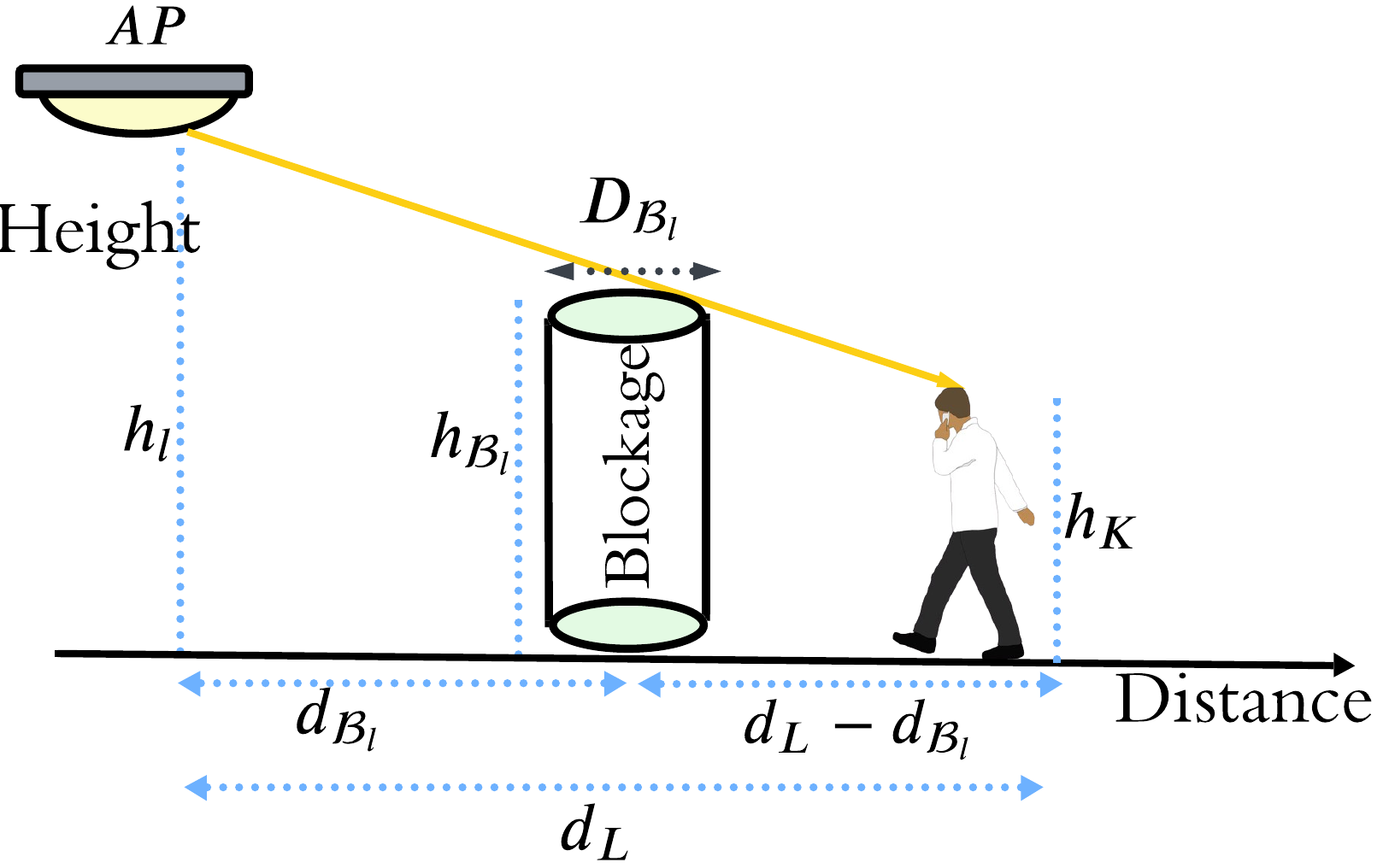}
    \caption{Blockage scenario.}
    \label{fig:Blockage}
    \vspace{-17pt}
\end{figure}

\vspace{-7pt}
\noindent where \eqref{eq:cons1} is the QoS constraint, and  \eqref{cons:2} and \eqref{cons:3} control the IRS mirror rotation angles. 
The optimization problem in \eqref{eq:P} is non-convex due to the effect of the constraints in \eqref{cons:2} and \eqref{cons:3} on the channel gain. Deterministic algorithms can be used to solve such problems at a complexity of $ \mathcal{O} (K+2(M_{ij}))^{3}$ at each iteration. Moreover, they operate at low accuracy in real-time and highly dynamic scenarios where user locations and traffic demand might change frequently.
Therefore, we propose a DRL algorithm to solve the optimization problem in practical systems.

\section{Intelligent RL solution based on DRL}\label{Sec:RL}

The optimization problem  \textbf{P} is reformulated as MDP model to enable the RL agent to discover an optimal policy by interacting with the environment. MDP provides a mathematical framework for modeling decision-making scenarios, defined by five essential components: $(\mathcal{S}, \mathcal{A}, \mathcal{R}, \mathcal{P}, \gamma)$, where $\mathcal{S}$ denotes the state space, $\mathcal{A}$ represents the action space, $\mathcal{R}$ defines the reward space, $\mathcal{P}$ captures transition probabilities, and $\gamma \in (0,1)$ is the discount rate. At each timestep $t$, the agent observes state $s_t$, $ s_t\in \mathcal{S}$, and takes action $a_t$, $a_t\in \mathcal{A}$, according to its policy. The environment then transitions to a new state $s_{t+1}$ with probability $P(s_{t+1}, r_t|s_t, a_t)$  \cite{sutton2018reinforcement}. 

 In our model, $\mathcal{S}$ is a set of states, and at each timestep $t$, the agent receives state $s_t$ from the environment consisting of the user's positions, channel gains, the current IRS rotation angles, and the minimum QoS of each user.


Furthermore, our agent takes decisions based on the received state from the action space $\mathcal{A}$. From \eqref{eq:P}, the action at the time step $t$ is given by the angles of the rotational mirrors, $\boldsymbol{\vartheta}_{m,t}$, and $\boldsymbol{\varphi}_{m,t}$. This enables the IRS to control and steer the signal towards users, maintaining connectivity in both mobility and blockage scenarios. For taking action $a_t$ in state $s_t$, the agent receives an immediate reward, $r_t$, and moves to the next state $s_{t+1}$. 
 
The reward in our model must reflect 
the objective of \eqref{eq:P} . That is 
\vspace{-7pt}
    \begin{equation} 
        r_t = {R_k} - \lambda_1 \cdot {QoS_k},
        \label{eq:reward}
   \end{equation}
\noindent where $\lambda_1$ is the penalty for not satisfying the target data rate for user $k$. This penalty function is an additional measure to prevent leaving any users unsatisfied while achieving a positive reward. The penalty function is defined as
    \begin{equation}
    \lambda_{1} = \begin{cases}
    0, & \text{if } \quad R_k \leq R_{\text{min},k}, \\
    \\1, & \text{if } \quad R_k \geq R_{\text{min},k},
    \end{cases}
    \end{equation}
To solve the formulated MDP problem, we use DRL, which is a model-free algorithm designed for continuous action spaces using deterministic policy gradients and DQL techniques. The DRL architecture employs an actor-critic framework to solve the challenge of high-dimensional control tasks. 
In our context, the actor, $\mu(s | \theta^{\mu})$, maps state $s$ to action $a$ to output continuous rotations of IRS angles. The actor-network is trained to maximize the expected reward by \cite{tan2021reinforcement}

\begin{equation}
\nabla_{\theta^{\mu}} J \approx \mathbb{E} \left[ \nabla_a Q(s, a | \theta^Q) \big|_{a = \mu(s)} \nabla_{\theta^{\mu}} \mu(s | \theta^{\mu}) \right],
\end{equation}
This gradient ascent ensures that the IRS configuration adapts to dynamic channel conditions. On the other hand, the critic-network, $Q(s, a | \theta^Q)$,  estimates the action-value function $Q(s,a)$, and is updated by minimizing the Bellman error as
\begin{equation}
L = \mathbb{E} \left[ \left( r + \gamma Q(s', \mu(s' | \theta^{\mu'}) | \theta^{Q'}) - Q(s, a | \theta^Q) \right)^2 \right].
\end{equation}

\noindent where $s'$ is the next state, $\gamma \in (0,1)$ is the discount factor, and  $\mu'$ and $Q'$ are the actor and critic target networks, respectively. The actor and critic update their parameters to obtain the policy that maximizes the reward.
\begin{algorithm}[H]
\caption{The proposed DRL algorithm for IRS-assisted VLC networks}
\label{algo}
\begin{algorithmic}
\State \textbf{Initialize:}
Set $t$ = 0, Actor-network $\mu(s|\theta^\mu)$ with random weights $\theta^\mu$, 
 Critic network $Q(s,a|\theta^Q)$ with random weights $\theta^Q$,
 Target networks $\mu'$ and $Q'$ with weights $\theta^{\mu'} \leftarrow \theta^\mu$, $\theta^{Q'} \leftarrow \theta^Q$, 
 Replay buffer $\mathcal{D}$,
 Indoor IRS VLC environment parameters.

\State Obtain the initial state $s_0$ of the environment.

\For{episode = 1 to Episodes}
   \For{step $t$ = 1 to $\mathcal{T}$}
       \State Select action $a_t = \mu(s_t|\theta^\mu) + \mathcal{N}_t$ 
       \State Execute $a_t$, compute the immediate reward $r_t$ by \eqref{eq:reward} and observe the new state $s_{t+1}$
       \State Store transition $(s_t, a_t, r_t, s_{t+1})$ in $\mathcal{D}$
       \State Sample minibatch of ${\mathcal{N}_b}$ transitions 
       
       $(s_i, a_i, r_i,s_{i+1})$ from $\mathcal{D}$
       \State Set $y_i = r_i + \gamma Q'(s_{i+1}, \mu'(s_{i+1}|\theta^{\mu'})|\theta^{Q'})$
       \State Update critic by minimizing: 
       
       $L = \frac{1}{\mathcal{N}_b}\sum_i(y_i - Q(s_i,a_i|\theta^Q))^2$
       \State Update actor using policy gradient: 
       
       $\nabla_{\theta^\mu}J \approx \frac{1}{\mathcal{N}_b}\sum_i\nabla_aQ(s_i,a_i|\theta^Q)|_{a_i=\mu(s_i)} \nabla_{\theta^\mu}\mu(s_i|\theta^\mu)$
       \State Update target networks:
       \State $\theta^{Q'} \leftarrow \tau\theta^Q + (1-\tau)\theta^{Q'}$
       \State $\theta^{\mu'} \leftarrow \tau\theta^\mu + (1-\tau)\theta^{\mu'}$
       \State Apply penalties if any of the constraints in \eqref{eq:P} are violated;
       \State update state $s_t \leftarrow s_{t+1}$
   \EndFor
   \If {the constraints in \eqref{eq:P} are met or max timesteps reached}
   \State End episode
\EndIf
\EndFor
\end{algorithmic}
\end{algorithm}


For better understanding, our DRL algorithm training is provided in Algorithm \ref{algo}. The actor neural network produces the actions for IRS orientation, and the critic neural network gives an evaluation of the actions. In each training episode, the algorithm is adaptive to the user positions and minimum user rate requirements. First, we initialize the DRL architecture along with IRS angles before moving forward to the optimization process. At each time step $t$, the current system state $s_t$ is observed, actions $a_t$ are taken and implemented in the environment, and the decision is evaluated to get the reward $r_t$ that is received by the agent with the new state $s_{t+1}$. The training process is based on experience replay buffer $\mathcal{D}$, where all the experience transitions are stored in the replay buffer and then sampled randomly in mini-batches $\mathcal{N}_b$ to update both actor $\mu$ and critic networks $Q$. The critic network is updated in order to minimize the difference error, and the actor-network is updated using the policy gradient method. Both target networks $\mu'$ and $Q'$ are updated by the soft update method to train the network to increase stability. By considering the constraints in (\ref{eq:cons1}), (\ref{cons:2}), and (\ref{cons:3}), this approach allows the system to adaptively optimize IRS orientation every time user locations and user demands change, thereby improving the system performance in real-time.
 

\begin{figure}[t]
\begin{center}\hspace*{0cm}
\includegraphics[width=1\linewidth,height=7.1cm,width=9.1cm]{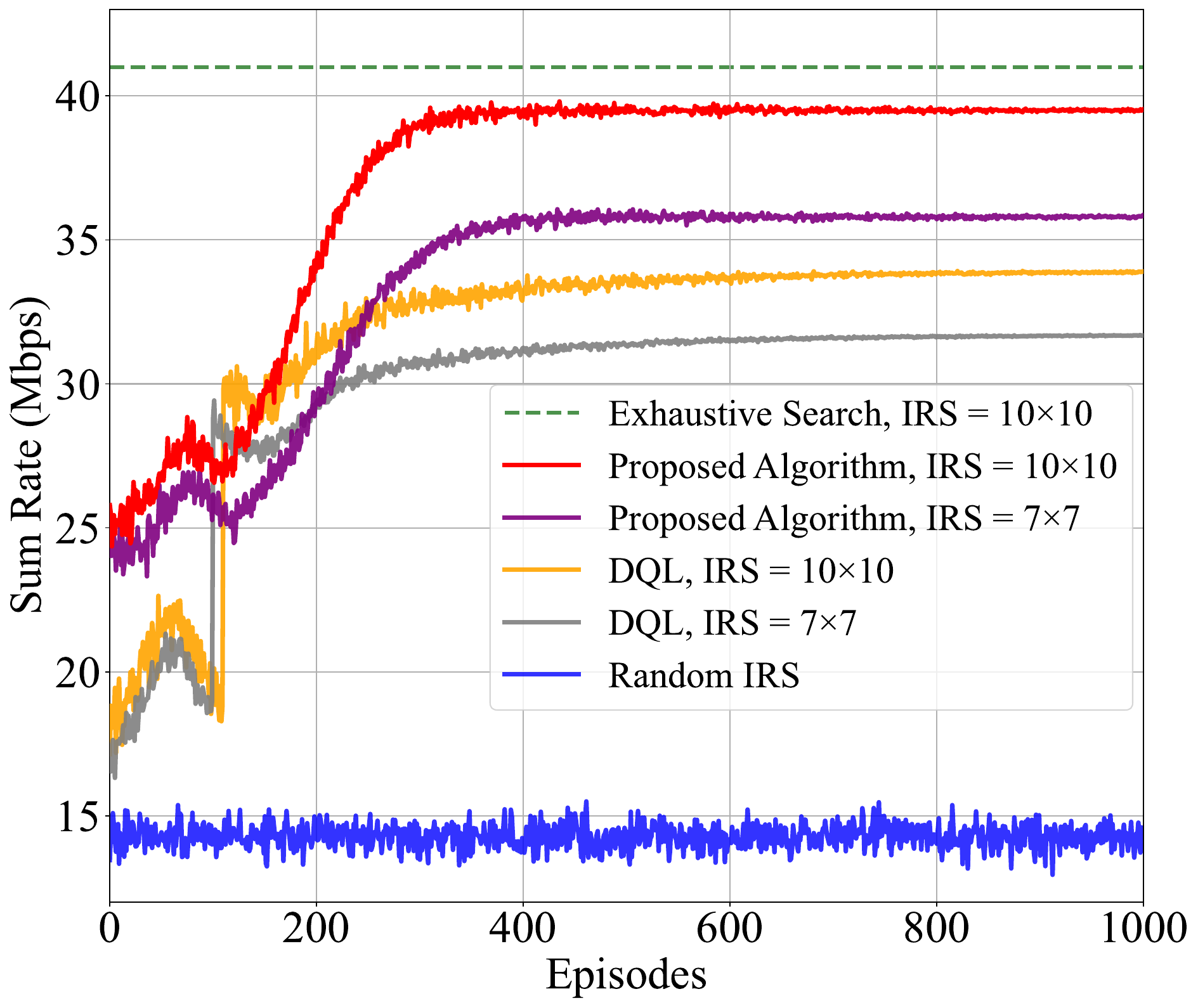}
\end{center}
\caption{Sum rates versus iterations. $P=2$ W.}\label{fig:Fig4}
\end{figure}


\section{performance evaluation}
\label{Sec:Results}

\subsection{ Simulation configuration}    
To evaluate the performance of our proposed algorithm, we consider the system model represented in Section \ref{Sec:System_model} in an indoor environment with 5$m$ $\times$ 5$m$ $\times$ 3$m$ dimensions. An LED-based AP is placed in the centre of the ceiling with a half-power semi-angle equal to $60^0$. An IRS mirror array is mounted on one wall. The mirror reflectivity is equal to 0.95, and each mirror has an effective area of 25 $cm$ $\times$ 10 $cm$ unless specified otherwise. On the communication plane, five active users (\(K=5\)) are distributed with different data rate requirements and a minimum data rate set to $R_{min}$ = 1 Mbps per user. Each user moves within the room at a randomly chosen speed [0,2] m/s, and the pause period is set to zero for the sake of simplicity. Other system parameters are shown in Table \ref{tab:sim_parameters}.

The simulation is implemented using Python 3.10. Our algorithm was trained over 1000 episodes. The hyperparameters are set carefully to prevent overfitting during training. We set the learning rate $\alpha= 0.05$, discount factor $\gamma = 0.9$, target network update $\tau = 0.01$, max buffer size $\mathcal{D}$ = 10000, and the mini batch size $\mathcal{N}_b$= 32. The exploration strategy is $\epsilon$ greedy, with an initial exploration rate of 0.995, which decreases to 0.0001. This helps the network to first explore the state space and then exploit the learned behaviors. The number of hidden layers is 2 for the actor and critic networks. The rectified linear unit (ReLU) activation function is used across all hidden layers, while Tanh is used for the actor-network. The network is optimized using the adaptive moment estimation (Adam) optimizer. The actor-network consists of 2 hidden layers with 128 and 64 neurons, respectively, while the two hidden layers of the critic network have 256 and 128 neurons, respectively. The actor-network and critic-network are trained with learning rates of $1e^{-3}$ and $1e^{-2}$, respectively. 


\begin{figure}[t]

\begin{center}\hspace*{0cm}
\includegraphics[width=1\linewidth,height=7.1cm,width=9cm]{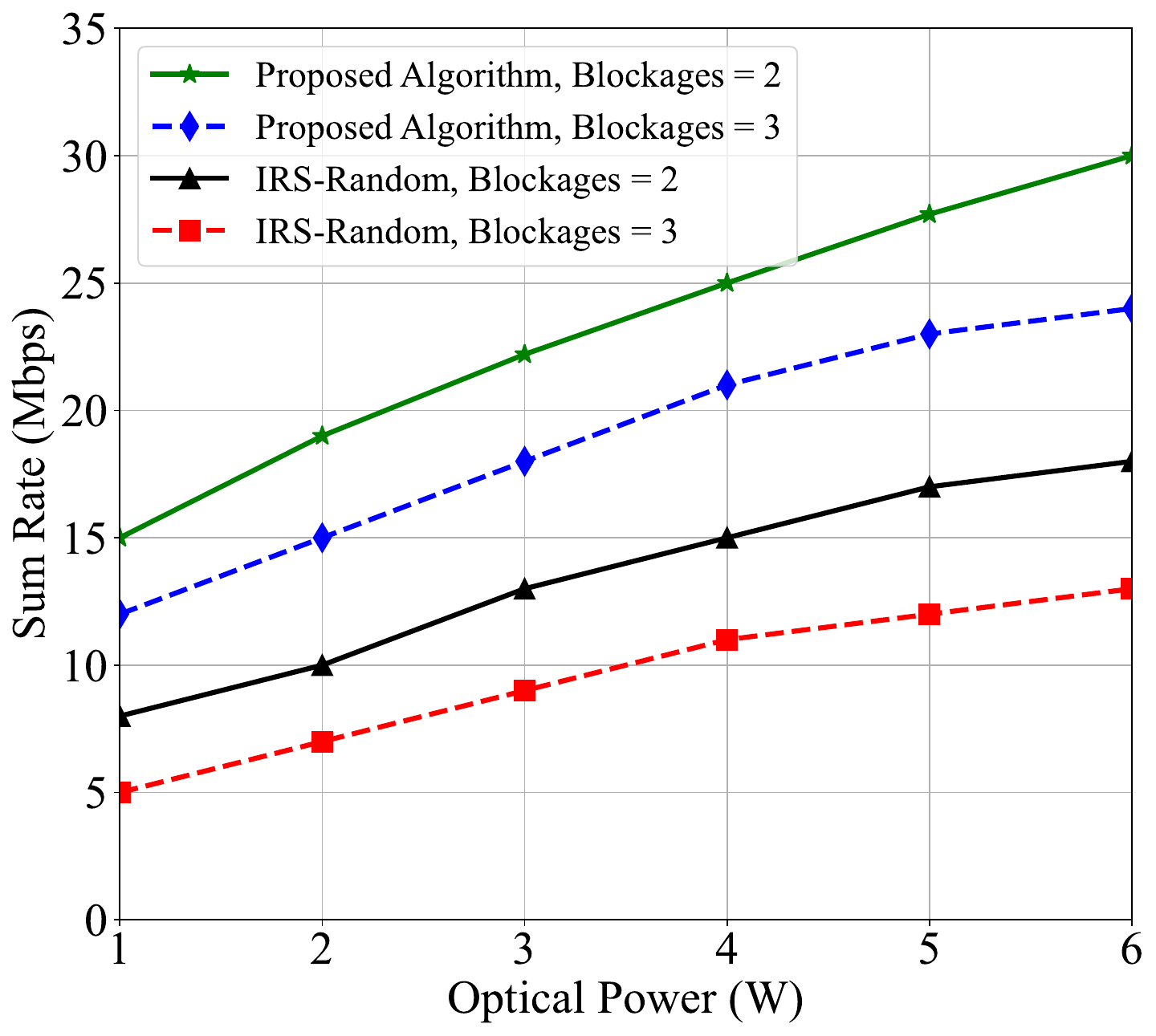}
\end{center}
\caption{Sum rates versus the transmitted optical power considering different numbers of blockages. IRS = 100.}\label{fig:Fig5}
\vspace{-3mm}
\end{figure}


\begin{table}[!t]
    \caption{Simulation Parameters} 
    \centering
    \begin{tabular}{|l | c|}
        \hline
        \textbf{Parameter} & \textbf{Value} \\
        \hline
        Photodetector physical area & 20 $mm^2$ \\
        \hline
        Concentrator refractive index & 1.5 \\ 
        \hline
        Optical filter gain & 1 \\
           \hline
        Photodetector responsivity & 0.4 \\
            \hline
          Photodetector bandwidth &  $20$ MHz\\
        \hline
        Number of IRS mirrors & 7 $\times$ 7, 10 $\times$ 10 \\
           \hline
           IRS reflection coefficient & 0.95\\
            \hline
    \end{tabular}   
    \label{tab:sim_parameters}
\end{table}
\subsection{ Simulation Results}

To verify the performance of our proposed algorithm, we compare our training results with the exhaustive search algorithm, the traditional DQL algorithm, and random IRS orientations. Fig. \ref{fig:Fig4} shows sum rates versus iterations for both the proposed algorithm and DQL algorithm, considering different numbers of IRS elements.  It can be seen that the proposed algorithm outperforms the DQL algorithm and the random IRS in terms of sum rate since our proposed algorithm can learn and optimize continuous actions for IRS orientations. Note that, in DQL, the actions are determined by the quantization of the angle values. This limitation makes the DQL algorithm unable to place the reflected signal optimally for mobile users, and hence, results in lower sum rates compared to the DRL algorithm. The comparison results verify the efficiency of the proposed algorithm in the time-varying dynamic IRS-assisted VLC environment.
Moreover, the sum rate increases with the number of IRS elements when using our proposed algorithm compared to DQL. This is due to the actor-critic architecture, which allows it to provide more accurate and continuous control of the IRS elements than the discrete control mechanism in DQL. Therefore, the proposed algoritm can steer the signal towards mobile users more efficiently than DQL, which needs discretization to handle continuous actions. The training is performed offline before deployment. Once trained, the DRL policy enables real-time online inference in 0.5 ms per decision, making it faster than traditional optimization methods while maintaining near-optimal performance.

\begin{figure}[t]
\begin{center}\hspace*{0cm}
\includegraphics[width=1\linewidth,height=7.1cm,width=9cm]{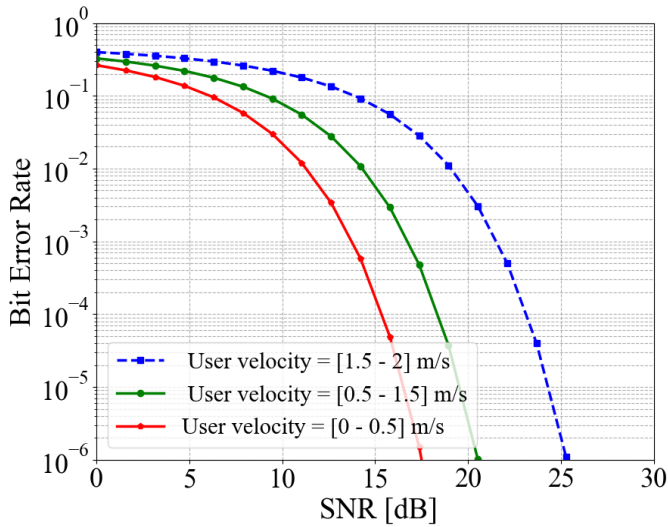}
\end{center}\caption{Bit error rates versus SNR.}\label{fig:Fig6}
\vspace{-7mm}
\end{figure}


Fig. \ref{fig:Fig5} shows the sum rate versus the optical transmitted power in different blockage scenarios. Our DRL algorithm is compared to a fixed orientation IRS as a benchmark scheme, where each IRS mirror points to a random direction. In other words, the IRS orientation in the benchmark scheme remains static throughout the serving time, regardless of user distributions and demands. It can be seen that the proposed DRL algorithm achieves up to 48\% higher sum rates compared to the benchmark scheme when two blockages and moderate mobility (i.e., user velocity $\approx$ 1 m/s) are considered. This is due to the ability of the DRL algorithm to steer the mirrors every time the users move and/or encounter blockages. For instance,  when the transmitted optical power is equal to 2 W, the proposed algorithm achieves almost 19.2 Mbps compared to 10 Mbps achieved by the benchmark scheme in the scenario of two blockages.

Fig. \ref{fig:Fig6} shows the bit error rate (BER) versus the SNR when IRS$ =10 \times 10$ are deployed, and different user velocities are considered. It can be seen that the BER increases with the speed of the user. For instance, when the SNR is  15 dB, the BER  increases from $10^{-4}$ to $10^{-1}$. It is worth mentioning that as the user moves faster, it becomes more challenging for our algorithm to precisely track its location and steer the mirrors towards it. However, as the SNR increases, the proposed algorithm achieves a BER below $10^{-3}$.

\section{Conclusion} \label{Sec:Conc}
This paper explored the application of mirror-based IRS to improve signal coverage and mitigate LoS blockages in dynamic indoor VLC systems. We addressed the challenge of optimizing IRS mirror orientation by formulating the problem as a MDP and applying the DRL algorithm, a powerful DRL technique. The proposed DRL-based solution enables real-time adaptability to user mobility and environmental blockages. Simulation results validate the effectiveness of our approach, showing that it consistently outperforms conventional DRL methods and yields significant gains in sum rate over random-orientation IRS configurations. 

\begingroup
\small
\bibliographystyle{IEEEtran}  
\bibliography{References}
\endgroup
\end{document}